\documentstyle[epsf]{article}

\newtheorem{Def}{Definition}
\newtheorem{Theo}{Theorem}
\newtheorem{Rem}{Remark}
\newtheorem{Prop}{Proposition}
\newtheorem{Cor}{Corollary}

\newcommand{\ba}{{\bf a}}
\newcommand{\be}{{\bf e}}
\newcommand{\bx}{{\bf x}}
\newcommand{\bt}{{\bf t}}
\newcommand{\bell}{{\bf \ell}}
\newcommand{\by}{{\bf y}}
\newcommand{\D}{{\Delta}}

\newcommand{\EE}{{\bf E}}
\newcommand{\CC}{{\bf C}}
\newcommand{\ZZ}{{\bf Z}}
\newcommand{\RR}{{\bf R}}

\newcommand{\bX}{{\bf X}}
\newcommand{\bF}{{\bf F}}

\newcommand{\ra}{\rightarrow}
\newcommand{\bOm}{{\bf \Omega}}
\newcommand{\blambda}{{\bf \lambda}}

\newcommand{\bPhi}{{\bf \Phi}}

\newcommand{\calC}{{\cal C}}

\newcommand{\cF}{{\cal F}}
\newcommand{\cL}{{\cal L}}
\newcommand{\cP}{{\cal P}}
\newcommand{\cQ}{{\cal Q}}
\newcommand{\cR}{{\cal R}}

\newcommand{\gl}{{\bf l}}

\newcommand{\gt}{{\bf t}}

\begin{document}

\begin{center}
{\Large \sc Integrable Discrete Geometry:} 

{\Large \sc the Quadrilateral Lattice,} 

{\Large \sc its Transformations and Reductions}
\bigskip

{\large Adam Doliwa$^{1,\dagger}$, Paolo Maria Santini$^{2,3,\S}$}

\bigskip

{\it $^1$Instytut Fizyki Teoretycznej, Uniwersytet Warszawski \\
ul. Ho\.{z}a 69, 00-681 Warszawa, Poland

\smallskip

$^2$Istituto Nazionale di Fisica Nucleare, Sezione di Roma\\
P.le Aldo Moro 2, I--00185 Roma, Italy

\smallskip

$^3$Dipartimento di Fisica, Universit\`a di Roma "La Sapienza"\\
P.le Aldo Moro 2, I--00185 Roma, Italy
}

\bigskip

$^\dagger$e-mail: {\tt Adam.Doliwa@fuw.edu.pl}

$^{\S}$e-mail:  {\tt Paolo.Santini@roma1.infn.it}

\bigskip

\end{center}

\begin{abstract}

\noindent 
We review recent results on Integrable Discrete Geometry. 
It turns out that most of the known (continuous and/or discrete) 
integrable systems are particular symmetries of the quadrilateral 
lattice, a multidimensional lattice characterized by the planarity 
of its elementary quadrilaterals. Therefore the linear property of 
planarity seems to be a basic geometric property underlying 
integrability. We solve the initial value problem for the quadrilateral
lattice and we present the geometric meaning of its $\tau$--function, as the
potential connecting its forward and backward data. We present the theory of
transformations of the quadrilateral lattice, which is based on the discrete
analogue of the theory of rectilinear congruences. In particular, we discuss the
discrete analogues of the Laplace, Combescure, L\'evy, radial and
fundamental transformations and their interrelations. We also 
show how the sequence of 
Laplace transformations of a quadrilateral surface is described by the
discrete Toda system in three dimensions. We finally show that these classical
transformations are strictly related to the basic operators associated with
the quantum field theoretical formulation of the multicomponent
Kadomtsev--Petviashvilii hierarchy. We review the properties of
quadrilateral hyperplane lattices, which play an interesting role in the
reduction theory, when the introduction of additional geometric
structures allows to establish a connection between point and hyperplane
lattices.  We present and fully characterize some geometrically 
distinguished reductions of the quadrilateral 
lattice, like the symmetric, circular and Egorov lattices; we review also
basic geometric results of the theory of quadrilateral lattices in quadrics,
and the corresponding analogue of the Ribaucour reduction of the 
fundamental transformation.

We finally remark that the equations characterizing the above
lattices are relevant in physics, being integrable
discretizations of equations arising in hydrodynamics and in quantum
field theory.

\end{abstract}

\section{Quadrilateral Lattices}

\subsection{The Quadrilateral Lattice}

Our first goal is to build a lattice which is integrable by 
construction. For the sake of simplicity, we consider first a 
2-dimensional lattice $\bx : \ZZ^2 \rightarrow \RR^M$. Given the three 
arbitrary points $\bx (n_1,n_2),~
\bx (n_1+1,n_2),~\bx (n_1,n_2+1)$, the forth point 
$\bx (n_1+1,n_2+1)$ is fixed prescribing a rule 
(the one defining the lattice);  
since we want an {\it integrable lattice}, this {\it rule must be 
linear}. We shall impose the simplest linear rule: {\it planarity}; 
i.e., the forth point $\bx (n_1+1,n_2+1)$ will belong to the plane 
generated by the first three points $\bx (n_1,n_2),~
\bx (n_1+1,n_2)$ and $\bx (n_1,n_2+1)$ or, in algebraic terms, 

\begin{equation}  \label{eq:Laplace12}
\D_1\D_2\bx=(T_{1} A_{12})\D_1\bx+
(T_2 A_{21})\D_2\bx ,
\end{equation}
where $T_i$ is the translation operator in the $i$ direction and 
$\D_i = T_i -1$ is the corresponding difference operator. Given two discrete 
initial curves (two sequences of points) $\{\bx^{(0)}_1\}$, $\{\bx^{(0)}_2\}$ 
and two arbitrary sets of functions $A_{12}(n_1,n_2),~ 
A_{21}(n_1,n_2): \ZZ^2 \rightarrow \RR$, the planarity constraint allows one 
to construct uniquely a two-dimensional lattice, which we have called 
{\it quadrilateral} (or {\it planar}). The quadrilateral surface was first 
proposed in \cite{Sauer} as the proper discrete analogue of a conjugate 
net on a surface, without any connection with integrability. 

Indeed, at the moment, the above construction does not 
seem to have anything to do with integrability; to show its profound 
connection with the familiar integrability schemes it is necessary to 
generalize the picture to higher dimensions, imposing the planarity 
constraint on each two - dimensional $(ij)$ surface of the lattice \cite{DS1}.

\begin{Def}
An $N$ dimensional lattice $\bx : \ZZ^N \rightarrow \RR^M$ is a 
quadrilateral lattice (QL) iff its elementary quadrilaterals 
$\{\bx ,T_i\bx , T_j\bx ,T_iT_j\bx \}$ are planar; i.e., iff       
\begin{equation}  \label{eq:Laplace}
\D_i\D_j\bx=(T_{i} A_{ij})\D_i\bx+
(T_j A_{ji})\D_j\bx ,\;\; i\ne j, \; \; \;  i,j=1 ,\dots, N.
\end{equation}

\end{Def}
The planarity constraints (\ref{eq:Laplace}) are compatible only for 
the special class of data $A_{ij}$ satisfying the nonlinear system

\begin{equation} \label{eq:MQL-A}
\D_k A_{ij} =
 (T_jA_{jk})A_{ij} +(T_k A_{kj})A_{ik} - (T_kA_{ij})A_{ik},
\;\; i\neq j\neq k\neq i.
\end{equation}
These equations characterize the quadrilateral lattice and we refer to them 
as the QL equations. They were first derived
in~\cite{BoKo} as integrable discrete analogues of the Darboux
equations for conjugate nets, but without any geometric
characterization.
 
We remark that, if a quadrilateral lattice  
exists, it is integrable by construction, since it is built 
out of the linear constraints (\ref{eq:Laplace}). Indeed, in the language 
of the 
theory of integrable systems, the planarity constraints   
correspond to the set of linear spectral problems (\ref{eq:Laplace}) and 
the resulting QL and its characterizing equations (\ref{eq:MQL-A}) 
correspond 
to the integrable nonlinear equations arising as the compatibility 
condition for 
such spectral problems. 

Actually, in connection with integrability, 
certain constructions based on the quadrilateral surfaces were already 
known: the
integrable discrete analogue of isothermic surfaces~\cite{BP2} and
the Laplace sequence of quadrilateral surfaces~\cite{DCN}, which
provides the geometric meaning for the Hirota equation (see also 
Section~\ref{sec:Hir} and~\cite{Dol-Hir} for more details). 

It is often convenient to reformulate  equations (\ref{eq:Laplace}) as 
a first order system \cite{DS1}. To do so we introduce the suitably scaled 
tangent vectors $\bX_i$, $i=1,...,N$:

\begin{equation}  \label{def:HX}
\D_i\bx = (T_iH_i) \bX_i,
\end{equation}
in such a way that the $j$-th variation of $\bX_i$ is proportional to $\bX_j$ 
only:

\begin{equation} \label{eq:lin-X}
\D_j\bX_i = (T_j Q_{ij})\bX_j,    \; \; \; i\ne j \; .
\end{equation}

\bigskip

\epsffile{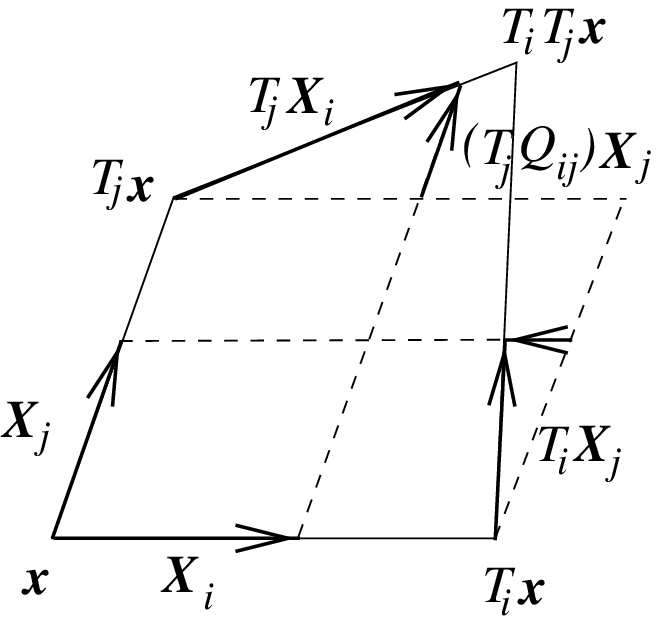}

Figure 1.

\bigskip

The compatibility condition for the system (\ref{eq:lin-X})
gives the following new form of the QL equations

\begin{equation} \label{eq:MQL-Q}
\D_kQ_{ij} = (T_kQ_{ik})Q_{kj}, \;\;\; i\neq j\neq k\neq i
\end{equation}
and the scaling factors $H_i$, called the Lam\'e coefficients, solve the 
linear equations

\begin{equation} \label{eq:lin-H}
\D_iH_j = (T_iH_i) Q_{ij}, \; \; \; i\ne j \; ,
\end{equation}
 whose compatibility gives equations (\ref{eq:MQL-Q}) again; moreover
$A_{ij}= \frac{\D_j H_i}{H_i}\; i\ne j$.

The solution of the initial value problem for the QL is contained in the 
following result \cite{DS1}.

\begin{Theo}[The Initial Value Problem for the QL]
Given $N$ initial discrete curves $\{\bx^{(0)}_i\}$, $i=1,..,N$, assigning the 
initial data $A^{(0)}_{ij}(n_i,n_j),A^{(0)}_{ji}(n_i,n_j)$,  
$i,j=1,..,N,~i\ne j$ (or, equivalently, the initial data 
$H^{(0)}_i(n_i),~Q^{(0)}_{ij},~Q^{(0)}_{ji}$), one constructs the initial 
$N(N-1)/2$ quadrilateral surfaces of the lattice applying equations 
(\ref{eq:Laplace}) (or, equivalently, equations (\ref{eq:lin-X})). 
Then the quadrilateral lattice $\bx$ follows uniquely from the planarity 
constraint.  
\end{Theo}

\bigskip

\epsffile{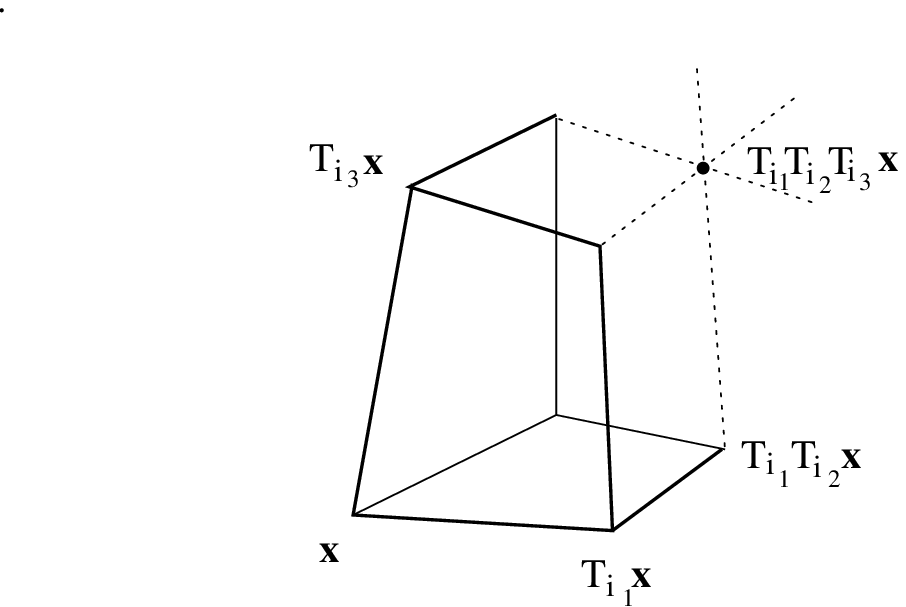}

Figure 2.

\bigskip

In the continuous limit:
\begin{equation}\label{eq:limit} 
\D_i  \sim \varepsilon \frac{\partial}{\partial u_i } =
 \varepsilon \partial_i \; ,~~Q_{ij}  \sim \varepsilon\beta_{ij},~~
  0 < \varepsilon < \! < 1
\end{equation}  
the QL reduces to an $N$ dimensional conjugate net in $\RR^M$,   
characterized by the famous Darboux equations~\cite{DarbouxOS}
\begin{equation} \label{eq:Darboux}
\partial_k \beta_{ij} = \beta_{ik}\beta_{kj}, \quad i\ne j \ne k \ne i.
\end{equation}

\subsection{The backward representation of the quadrilateral lattice and the 
geometric meaning of the $\tau$ function}

In this Section we define the backward data $\tilde{\bX}_i$, $\tilde{H}_i$, 
$\tilde{Q}_{ij}$ of the quadrilateral lattice. It turns out that the relation 
between the forward data $\bX_i$, $H_i$, $Q_{ij}$ and the backward data
is given in terms of the $\tau$--function \cite{DS4}, which is one of central
objects of the theory of integrable systems.

In the previous Section the quadrilateral lattice was built through a 
{\it forward construction}; it is of course possible to build the lattice 
also through a {\it backward construction}. The backward tangent vectors 
$\tilde{\bX}_i$ and the backward Lam\'e coefficients
$\tilde{H}_i$, $i=1,\dots,N$ are defined by the equations 
\begin{equation} \label{eq:b-H-X}
\tilde{\D}_i\bx = (T_i^{-1}\tilde{H}_i ) \tilde{\bX}_i \; , \qquad 
\hbox{or}
\quad \D_i\bx = \tilde{H}_i (T_i\tilde{\bX}_i ) \, ;
\end{equation}
in terms of the backward difference operator $\tilde{\D}_i := 1- T_i^{-1}$. 
The backward Lam\'e coefficients are again chosen in such a way 
that the $\tilde\Delta_i$ variation of $\tilde{\bX}_j$ is proportional to 
$\tilde{\bX}_i$ only. We define the backward rotation coefficients 
$\tilde{Q}_{ij}$ as the corresponding proportionality factors
\begin{equation} \label{eq:lin-bX}
\tilde\Delta_i\tilde{\bX}_j = (T_i^{-1} \tilde{Q}_{ij})\tilde{\bX}_i \; , 
\qquad \hbox{or}\quad \D_i\tilde{\bX}_j =  (T_i\tilde{\bX}_i)\tilde{Q}_{ij},    \quad i\ne j \; .
\end{equation}

\bigskip

\epsffile{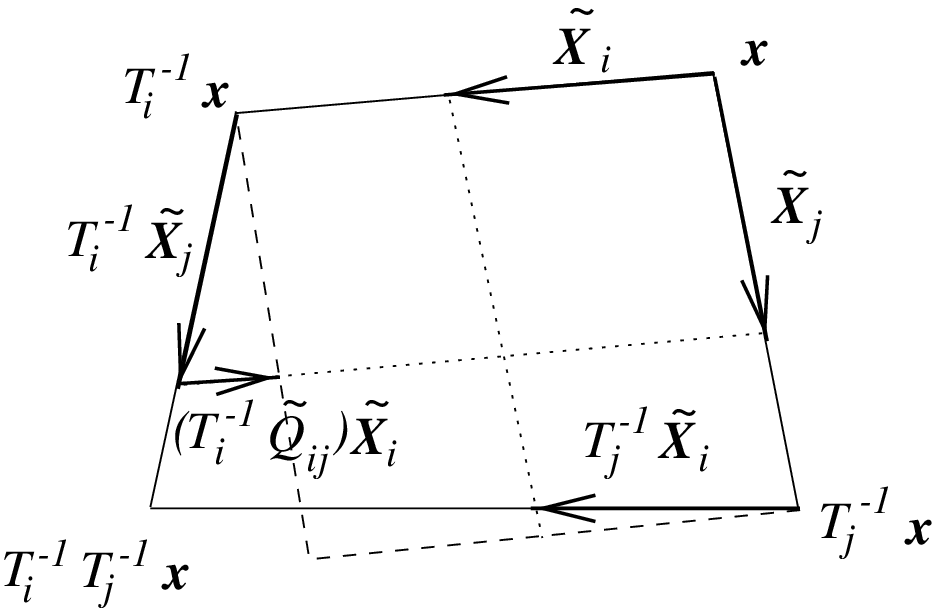}

Figure 3.

\bigskip

Comparing equations (\ref{eq:lin-H}) and (\ref{eq:lin-bX})
we see immediately that the new functions $\tilde{Q}_{ij}$ satisfy the MQL 
equations~(\ref{eq:MQL-Q}) as well. Moreover the new scaling factors 
$\tilde{H}_i$ satisfy the following system of linear equations
\begin{equation} \label{eq:lin-bH}
\D_j\tilde{H}_i =  (T_j\tilde{Q}_{ij})\tilde{H}_{j},   \; \; \; i\ne j \; ,
\end{equation}
whose compatibility condition gives again the QL equations 
(\ref{eq:MQL-Q}).

An easy consequence of equations (\ref{eq:b-H-X}), (\ref{eq:lin-bX}) and  
(\ref{eq:lin-bH}) is the following
\begin{Prop} 
The vector function $\bx: \ZZ^N \rightarrow \RR^M$ representing a
quadrilateral lattice satisfies the backward Laplace equation
\begin{equation} \label{eq:Laplace-b}
\tilde\D_i\tilde\D_j\bx = (T_i^{-1}\tilde{A}_{ij})\tilde\D_i\bx +
(T_j^{-1}\tilde{A}_{ji})\tilde\D_j\bx
\; , \quad i\ne j \; ,
\end{equation}
where, in the notation of this Section, 
$\tilde{A}_{ij} = \frac{\tilde\D_j \tilde{H}_i}{\tilde{H}_i}$.
\end{Prop}

The forward and backward rotation coefficients
$Q_{ij}$ and $\tilde{Q}_{ij}$ describe the same lattice $\bx$
from different points of view, therefore they are not independent.  
Indeed, defining the functions $\rho_i:\ZZ^N\to \RR$ as the 
proportionality factors between
$\bX_i$ and $T_i\tilde{\bX}_i$ (both vectors are proportional to $\D_i\bx$):
\begin{equation} \label{eq:def-rho}
\bX_i = - \rho_i ( T_i\tilde{\bX}_i) \; , \qquad T_iH_i = - 
\frac{1}{\rho_i}\tilde{H}_i \; , \quad
i=1,\dots ,N \; ,
\end{equation}
we have the following

\begin{Prop}
The forward and backward data of the lattice $\bx$ are related 
through the following formulas
\begin{equation} \label{eq:Q-Qt}
\rho_j T_j\tilde{Q}_{ij} =  \rho_i T_iQ_{ji} \; ,
\end{equation}
and the factors $\rho_i$ are first potentials satisfying
equations
\begin{equation} \label{eq:rho-constr}
\frac{T_j\rho_i}{\rho_i} = 1 - (T_iQ_{ji})(T_jQ_{ij}) \; , i\ne j \; .
\end{equation}
\end{Prop}

\begin{Rem}
Since $Q_{ij}$ and $\tilde{Q}_{ij}$ are both solutions of the QL
equations~(\ref{eq:MQL-Q}), then equations~(\ref{eq:Q-Qt}),
(\ref{eq:rho-constr}) describe a special symmetry transformation
of equations~(\ref{eq:Q-Qt}).
\end{Rem}

The RHS of equation~(\ref{eq:rho-constr}) is symmetric with
respect to the interchange of $i$ and $j$; this implies the
existence of a potential $\tau:\ZZ^N\to \RR$, such that
\begin{equation}
\rho_i = \frac{T_i\tau}{\tau} \; \; ;
\end{equation}
therefore equation~(\ref{eq:rho-constr}) defines the second potential $\tau$:
\begin{equation} \label{eq:tau}
\frac{(T_i T_j\tau)\tau}{(T_i \tau)(T_j\tau)} = 1 - (T_iQ_{ji})(T_jQ_{ij})
\; , \quad i\ne j \; .
\end{equation}
The potential $\tau$ connecting the forward and backward data:
\begin{eqnarray}
T_j(\tau\tilde{Q}_{ij})  & = & T_i(\tau Q_{ji}) \; , 
\label{eq:Q-Qt-tau}\\  T_i(\tau\tilde{\bX}_i) & = & \tau\bX_i \; , \\
\tau \tilde{H}_i & = & T_i(\tau H_i) \; ,
\end{eqnarray}
is the famous $\tau$-{\it function} of the quadrilateral lattice. In terms 
of the $\tau$ function, the QL equations read as follows \cite{DMMMS}:
\begin{equation} \label{eq:Hir-ij}
(T_i T_j \tau ) \tau = (T_i\tau) T_j\tau - (T_i\tau_{ji}) T_j\tau_{ij} \; ,
\end{equation}
\begin{equation} \label{eq:Hir-ijk}
(T_k \tau_{ij}) \tau  = (T_k\tau) \tau_{ij} + (T_k\tau_{ik}) \tau_{kj} \; ,
\end{equation}
where 
\begin{equation}
\tau_{ij} = \tau Q_{ij} .
\end{equation}

\section{Transformations of quadrilateral lattices}

We present the basic ideas and results of the theory of the Darboux type
transformations of the multidimensional quadrilateral lattice. Our approach
follows the spirit of the book~\cite{Eisenhart-TS}, which summarizes the theory of
transformations of (two dimensional) conjugate nets. We also name the
transformations of the quadrilateral lattice according to the
classical geometric terminology of the transformations of conjugate nets.

All the results of this section can be found in~\cite{TQL}, to whom we also refer
for proofs and more detailed explanation.

\subsection{Transformations and congruences}

To define the transformations we need another geometric object -- the {\it
congruence} -- which serves as the link between two quadrilateral lattices.

\begin{Def} \label{def:int-congr}
An $N$-dimensional {\it rectilinear congruence}
(or, simply, congruence) is a mapping $\gl : \ZZ^N \ra Gr_A(2,M+1)$
from the integer lattice to the
space of lines in $\RR^M$ such that every two neighboring lines
$\gl$ and $T_i \gl$, $i=1,...,N$, are coplanar.
\end{Def}

We remark that with any $N$ dimensional
quadrilateral lattices we may naturally associate $N$ congruences
as follows.

\begin{Def} \label{def:tan-congruence}
Given an $N$-dimensional quadrilateral lattice $\bx$,
its $i$-th {\it tangent congruence} $\gt_i(\bx)$ consists of the lines
passing through the points $\bx$ of the lattice and directed along
the tangent vectors $\D_i\bx$.
\end{Def}

Reversing that process, one can associate with any $N$-dimensional 
congruence $N$ lattices which are, in general, quadrilateral lattices.

\begin{Def}
The $i$-th {\it focal lattice} $\by_i(\gl)$ of a congruence $\gl$
is the lattice constructed out of the intersection points of the lines $
\gl$ with $T_i^{-1}\gl$.
\end{Def}

There exists a simple (but basic for the theory of transformations)
relation between quadrilateral lattices and congruences.

\begin{Def} \label{def:conj-congr-latt}
A quadrilateral lattice and a
congruence 
are called {\it conjugate} if there exists a one-to-one correspondence between
points of the lattice and lines of the congruence such that lines pass
through the corresponding points.
\end{Def}

\begin{Cor}
Focal lattices of congruences conjugate to quadrilateral lattices
are quadrilateral lattices.
\end{Cor}

The above introduced notions allow to define all the transformations
of quadrilateral lattices.

\begin{Def} \label{def:fundamental}
Two quadrilateral lattices are related
by the {\it fundamental transformation $\cF$ of Jonas} when they are conjugate 
to the same
congruence, which is called the congruence of the transformation.
\end{Def}

Usually one assumes that, in the fundamental transfomation, the congruence
is {\it transversal} to the lattice, i.e., non-tangent to the lattice in
corresponding points. Otherwise we get the following
reductions of the fundamental transformation
\begin{itemize}
\item the congruence of the transformation is the $i$-th tangent
congruence of the original lattice -- the {\it L\'evy transformation}
$\cL_i$
\item the congruence of the transformation is the $i$-th tangent
congruence of the new lattice -- the {\it adjoint
L\'evy transformation} $\cL_i^*$
\item both lattices are focal lattices of the congruence of the
transformation -- the {\it Laplace transformation}
\end{itemize}

\bigskip

\leavevmode\epsfxsize=7cm\epsffile{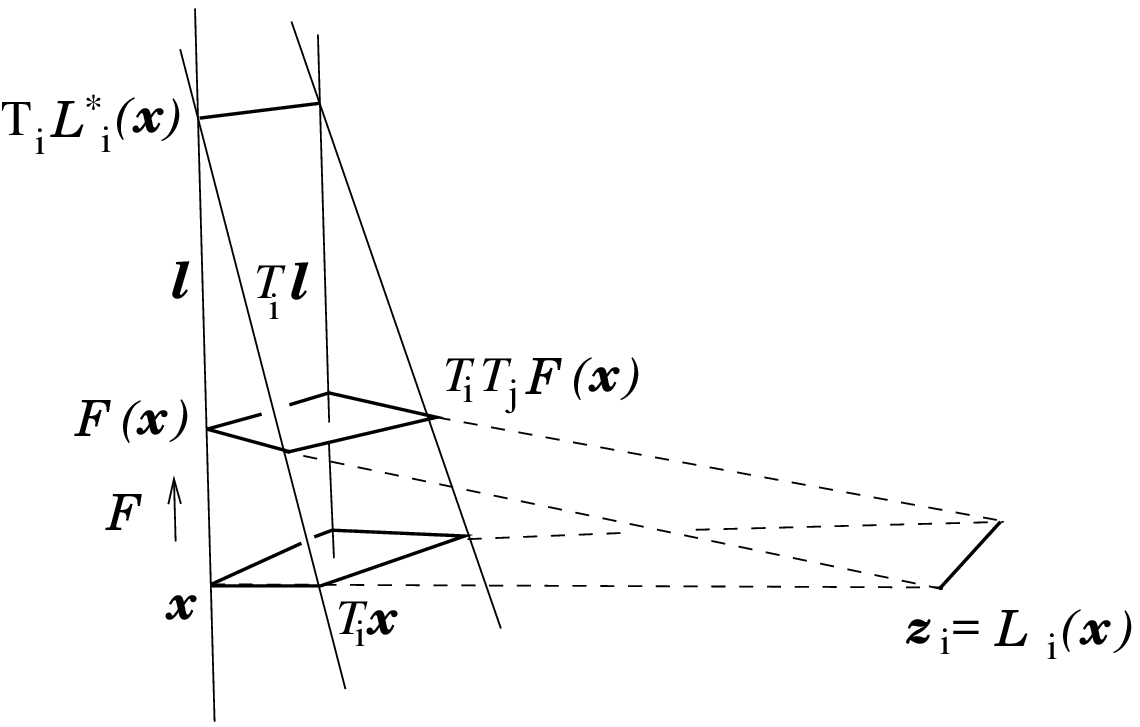}

Figure 4.

\bigskip

In addition, there are two useful special transformations
\begin{itemize}
\item the {\it projective} or {\it radial transformation}, when all the lines
of the congruence meet in a single point
\item the {\it Combescure transformation}, when the transformed lattice is
parallel to the original one
\end{itemize}

Let us present some formulas for the transformations, the details can be
found in~\cite{TQL}.

\begin{Theo} \label{th:fund}
The fundamental transformation $\cF(\bx)$ of the quadrilateral lattice $\bx$
is given by
\begin{equation} \label{eq:fund}
\cF(\bx) = \bx - \frac{\phi}{\phi_\calC}\bx_\calC \; \; ,
\end{equation}
where\\
i) $\phi:\ZZ^N\ra\RR$ is a new solution of the Laplace equation (\ref{eq:Laplace})
of the lattice $\bx$\\
ii) $\bx_\calC$ is the Combescure transformation vector, which is a solution
of the equations
\begin{equation} \label{eq:DixC} \D_i\bx_\calC = (T_i\sigma_i)\D_i\bx \; \; ,
\end{equation}
where, due to the compatibility of the system (\ref{eq:DixC}),
the functions $\sigma_i$ satisfy
\begin{equation} \label{eq:Djsi}
\D_j\sigma_i = A_{ij}(T_j\sigma_j - T_j\sigma_i) \; , \quad i\neq j \; ;
\end{equation}
moreover\\
iii) $\phi_\calC$ is a solution, corresponding to $\phi$, of the Laplace equation
of the lattice $\bx_\calC$, i.e.
\begin{equation}
\label{eq:DiphC} \D_i\phi_\calC = (T_i\sigma_i)\D_i\phi \; \; .
\end{equation}
\end{Theo}
The Combescure transformation vector $\bx_\calC$ is used to construct the
congruence of the transformation, and the function $\phi$ serves to place
points of the new lattice on the lines of the congruence. The definitions 
(\ref{eq:fund}) and (\ref{eq:DixC})-(\ref{eq:Djsi}) of the fundamental and
Combescure transformations were independently presented in~\cite{KoSchief2}.

\begin{Rem}
Notice that, given $\bx_\calC$ and $\phi$, then equation~(\ref{eq:DiphC})
determines $\phi_\calC$ uniquely, up to a constant of integration.
\end{Rem}
\begin{Cor}
One can stop the transformation of the lattice $\bx$ at the intermediate
level of construction of the vector $\bx_\calC$ to obtain the Combescure
transformation
\begin{equation}
\calC(\bx) = \bx + \bx_\calC \; ,
\end{equation}
characterized by the property that the tangent lines to both lattices in
corresponding points are parallel. 
\end{Cor}

\begin{Def} \label{def:Levy}
The $i$-th {\it L\'evy transform} $\cL_i(\bx)$ of the quadrilateral lattice
$\bx$ is a quadrilateral lattice
conjugate to the $i$-th tangent congruence of $\bx$.
\end{Def}
Since the congruence of the transformation is known, to construct the
transformation we need only the solution $\phi$ of the Laplace equation 
of the lattice $\bx$.
\begin{Prop} \label{prop:Levy}
The L\'evy transform $\cL_i(\bx)$ of the quadrilateral
lattice $\bx$ is given by
\begin{equation} \label{eq:Levy-transf}
\cL_i(\bx) = \bx - \frac{\phi}{\D_i\phi}\D_i\bx.
\end{equation}
\end{Prop}
Another way to reduce the fundamental transformation is to fix the way we
place the new lattice on the congruence.
\begin{Def} \label{def:adj-Levy} The $i$-th {\it adjoint L\'evy transform}
$\cL^*_i(\bx)$ of the quadrilateral lattice
$\bx$ is the $i$-th focal lattice of a congruence conjugate to $\bx$.
\end{Def}
To find the adjoint L\'evy transformation we need only the congruence
of the transformation, which can be find via the Combescure transformation
vector $\bx_\calC$. 
\begin{Prop}
The adjoint L\'evy transform of the lattice $\bx$ is given by
\begin{equation}  \label{eq:adj-Levy1}
\cL_i^*(\bx) = \bx - \frac{1}{\sigma_i}\bx_\calC \; ,
\end{equation}
where the functions $\sigma_i$ are defined in Theorem~\ref{th:fund}. 
\end{Prop}
The deepest reduction of the fundamental transformation is the Laplace
transformation where both the congruence of the transformation and the way
to place the new lattice on the congruence are fixed.
\begin{Def}
The {\it Laplace transform} $\cL_{ij}(\bx)$ of the quadrilateral lattice $\bx$
is the $j$-th focal lattice of its $i$-th tangent congruence.
\end{Def}

\bigskip

\leavevmode\epsfxsize=9cm\epsffile{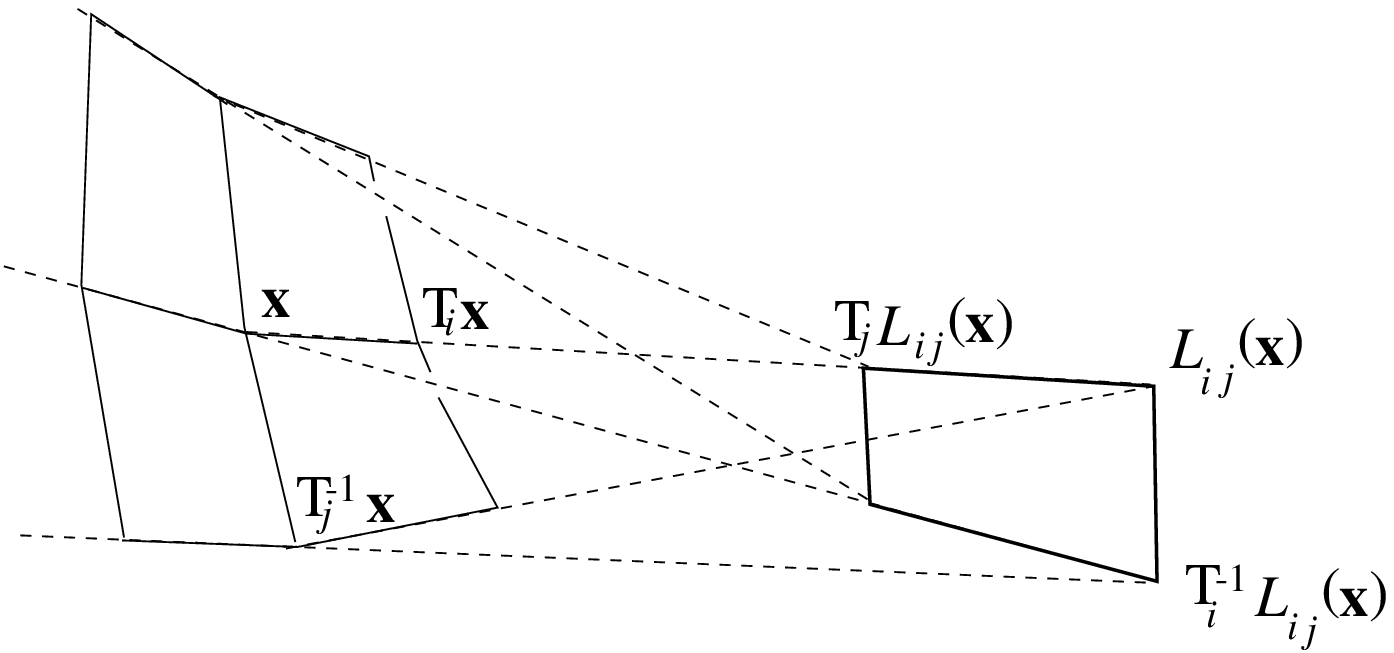}

Figure 5.

\bigskip

\begin{Prop} \label{prop:Laplace}
The Laplace transformation of the quadrilateral lattice $\bx$ is given by
\begin{equation} \label{def:Lij}
\cL_{ij}(\bx) = 
\bx - \frac{1}{A_{ji}} \D_i\bx.
\end{equation}
\end{Prop}
\begin{Cor}
The superpositions of Laplace transformations satisfy the following
identities
\begin{eqnarray} \label{eq:Lid-ij}
\cL_{ij} \circ \cL_{ji} &  = & {\rm id} \; , \\
\cL_{jk} \circ \cL_{ij} &= & \cL_{ik}, \label{eq:Lid-ijk1} \\
\cL_{ki} \circ \cL_{ij} &= & \cL_{kj}. \label{eq:Lid-ijk2}
\end{eqnarray}
\end{Cor}

Finally, the notion of radial (or projective) transformation is related 
to the so called radial congruence, whose focal lattices are 
degenerated to a single
point, which can be taken as the origin.
\begin{Def}
The radial (or projective)
transform $\cP(\bx)$ of the quadrilateral lattice $\bx$ is a
quadrilateral lattice conjugate to the radial congruence of 
$\bx$.
\end{Def}
\begin{Prop} \label{prop:radial}
The radial transform $\cP(\bx)$ is given by
\begin{equation} \label{eq:radial}
\cP(\bx) = \frac{1}{\phi}\bx,
\end{equation}
where $\phi:\ZZ^N\ra \RR$ is a solution of the Laplace equation
(\ref{eq:Laplace}) of the lattice $\bx$.
\end{Prop}
The degeneration of the fundamental transformation to the L\'evy, adjoint
L\'evy, the Combescure, radial, and the Laplace transformations can be shown
also on the analytic level. From another point of view, the fundamental
transformation can be decomposed into the superposition of the L\'evy and 
adjoint L\'evy transformations, or into the superposition of two Combescure and
one radial transformation. Moreover the fundamental transformation can be
put into the scheme of the Laplace transformation (preceeded by building a 
new
level on the original lattice).

\subsection{The Hirota equation and the Laplace transformations 
of quadrilateral surfaces}

\label{sec:Hir}

The Laplace transformations of two dimensional conjugate nets
generates the two dimensional Toda system~\cite{DarbouxIV,Mikhailov}.
Analogously, one can associate~\cite{DCN} (see also~\cite{Dol-Hir}
for more datails and new interpretations)
the integrable discrete version of the Toda system (the 
Hirota equation~\cite{Hir}) with the Laplace 
transformations of a two dimensional 
quadrilateral lattice. 

Given a quadrilateral surface $\bx:\ZZ^2\ra \RR^M$, consider the sequence
$\bx^{(k)}$ of quadrilateral surfaces obtained from $\bx$ 
by resursive application of the Laplace transformation $\cL_{12}$:
\begin{equation}
\bx^{(k)} = \left( \cL_{12}\right)^k (\bx) \; , \qquad k\in\ZZ \; .
\end{equation}
The above sequence, which is the proper analogue of the Laplace sequence
of conjugate nets, 
is well  defined due to formula~(\ref{eq:Lid-ij}).
The coefficients $A_{12}^{(k)}$, $A_{21}^{(k)}$ of the Laplace 
equations of the sequence of lattices satisfy the following system
\begin{eqnarray}
\label{eq:TodaA1}
\frac{A_{12}^{(k+1)}+ 1}{T_1A_{12}^{(k)}+1} & = & 
\frac{A_{21}^{(k)}}{T_2A_{21}^{(k)}} \; \; , \\
\label{eq:TodaA2}
\frac{A_{21}^{(k-1)}+ 1}{T_2A_{21}^{(k)}+1} & = & 
\frac{A_{12}^{(k)}}{T_1A_{12}^{(k)}} \; \; .
\end{eqnarray}

Notice that one can consider the parameter $k\in\ZZ$ along the Laplace 
sequence as the third discrete variable.  
Actually, the Laplace sequence of quadrilateral surfaces may be viewed 
as a deep reduction of the three dimensional quadrilateral lattice in 
the same sense like the Laplace transformations are reductions of the 
fundamental transformation. An elementary quadrilateral of such reduced 
(or degenerated) lattice with $T_3 = \cL_{12}$ 
is shown below.

\bigskip

\leavevmode\epsfxsize=8cm\epsffile{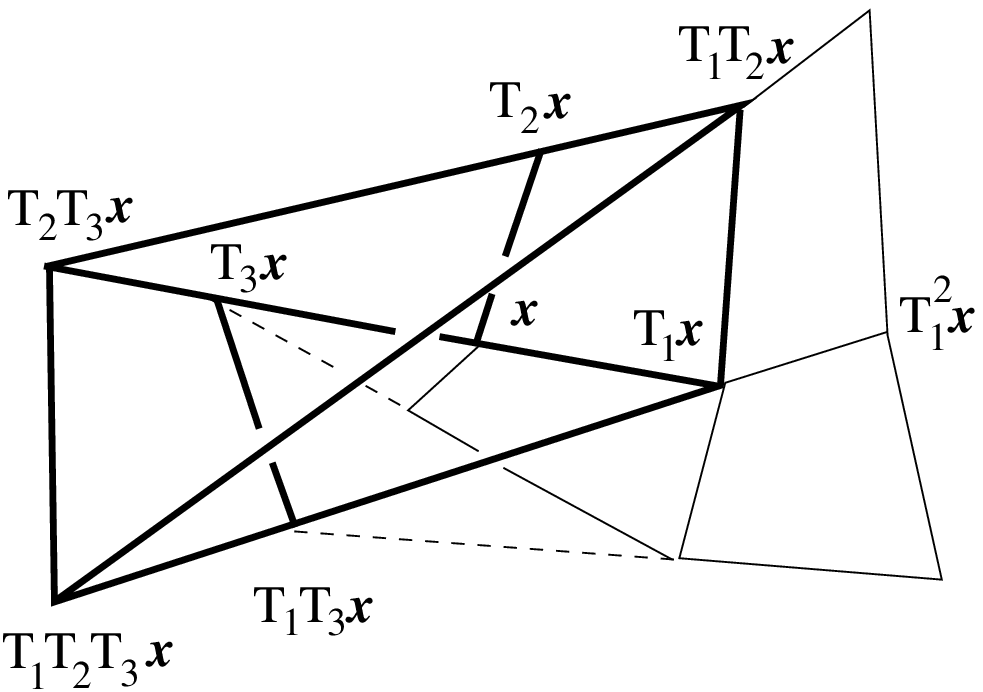}

Figure 6.

\bigskip

Define the function $K:\ZZ^2 \ra \RR$ as the cross-ratio of four collinear
points: $x$,
$\cL_{12}(x)$, $T_1 x$ and $T_j\cL_{12}(x)$. Simple derivation shows that
\begin{equation}
K = \frac{A_{21}(T_1A_{12} + 1) - T_2A_{21}}{(1+T_1A_{12})(1+ A_{21}} \; ,
\end{equation}
and equations (\ref{eq:TodaA1})-(\ref{eq:TodaA2})
imply that $K$ satisfies the gauge invariant
form of the Hirota equation 
\begin{equation} \label{eq:K}
T_2 \left( \frac{K^{(l+1)}+1}{K^{(l)}+1} \right)
T_1\left(\frac{K^{(l-1)}+1}{K^{(l)}+1} \right) = \frac{(T_1T_2
K^{(l)}) K^{(l)}}{(T_1K^{(l)})(T_2K^{(l)})} \; .
\end{equation}

\subsection{Superposition of fundamental transformations}
Consider $K\geq 1$ fundamental transformations $\cF_k(\bx)$, $k=1,...,K$,
of the quadrilateral lattice $\bx$, which are built from\\
i) $K$ solutions $\phi^k$, $k=1,...,K$ of the Laplace
equation of the lattice $\bx$;\\ 
ii) $K$ Combescure transformation vectors $\bx_{\calC,k}$, where
\begin{equation}
\D_i \bx_{\calC,k} = (T_i\sigma_{i,k})\D_i\bx \; , \; i=1,...,N \:,
\;k=1,...,K\; ,
\end{equation}
and $\sigma_{i,k}$ satisfy equations
\begin{equation} \label{eq:Djsik}
\D_j\sigma_{i,k} = A_{ij}(T_j\sigma_{j,k} - T_j\sigma_{i,k}) \; , 
\quad i\neq j \; ; 
\end{equation}
iii) $K$ functions $\phi^k_{\calC,k}$, which satisfy
\begin{equation}
\D_i \phi^k_{\calC,k} = (T_i\sigma_{i,k})\D_i\phi^k \; .
\end{equation}
We arrange functions $\phi^k$ in the $K$ component vector
$ \bPhi = ( \phi^1 ,\dots, \phi^K)^T$;
similarily, we arrange the
Combescure transformation vectors $\bx_{\calC,k}$ into the 
$M\times K$ matrix
$\bX_\calC = (\bx_{\calC,1}, ... ,\bx_{\calC,K}) $;
moreover we introduce\\
iv) the $K\times K$ matrix
$\bPhi_\calC= (\bPhi_{\calC,1},\dots,  \bPhi_{\calC,K})$, whose
columns are the $K$ component vectors
$\bPhi_{\calC,k} = (\phi^1_{\calC,k}, ..., \phi^K_{\calC,k})^T$,
which are the Combescure transforms of $\bPhi$
\begin{equation} \label{eq:int-phiC}
\D_i \bPhi_{\calC,k} = (T_i\sigma_{i,k})\D_i\bPhi \;  .
\end{equation}
\begin{Rem}
The diagonal part of $\bPhi_\calC$ is fixed by the initial
fundamental transformations. To find the off-diagonal part of $\bPhi_\calC$
we integrate equations~(\ref{eq:int-phiC}) introducing $K(K-1)$ arbitrary 
constants.
\end{Rem}
One can show that the vectorial fundamental transformation  
${\bF}(\bx)$
of the quadrilateral lattice $\bx$, which is defined as
\begin{equation} \label{eq:vect-fund}
{\bF}(\bx) = \bx - \bX_\calC \bPhi_\calC^{-1} \bPhi \;  ,
\end{equation}
is again a quadrilateral lattice. Moreover, the vectorial transformation
is the superposition of the fundamental transformations
\begin{equation}
{\bF}(\bx)= (\cF_{k_1}\circ \cF_{k_2}\circ\dots\circ\cF_{k_K} )(\bx) \; ,
\quad k_i \ne k_j \quad {\rm for} \quad i \ne j
\end{equation}
and does not depend on the order in which the transformations are taken.
In applying the fundamental transformations at the intermediate
stages, the transformation data should be suitably transformed as well.

To prove the
superposition and permutability statements, it is important to notice 
the following basic fact.
\begin{Theo}[Permutability theorem] \label{lem:sup-vect-fund}
Assume the following splitting of the data of the vectorial fundamental
transformation
\begin{equation}
\bPhi=\left( \begin{array}{c}\bPhi^{(1)} \\ \bPhi^{(2)} \end{array}
\right) \; , \quad
\bX_\calC = \left( \bX_{\calC(1)}, \bX_{\calC(2)} \right) \; , \quad
\bPhi_\calC = \left( \begin{array}{cc} \bPhi^{(1)}_{\calC(1)} &  
\bPhi^{(1)}_{\calC(2)} \\
\bPhi^{(2)}_{\calC(1)} &  \bPhi^{(2)}_{\calC(2)}  \end{array} \right) \; ,
\end{equation}
associated with the partition $K=K_1+K_2$. Then the vectorial fundamental
transformation ${\bF}(\bx)$ is equivalent to the following
superposition of vectorial fundamental transformations:\\
1. Transformation ${\bF}_{(1)}(\bx)$ with the data $\bPhi^{(1)}$,
$\bX_{\calC(1)}$, $\bPhi^{(1)}_{\calC(1)}$:
\begin{equation}
{\bF}_{(1)}(\bx) = \bx - \bX_{\calC(1)} \left( \bPhi^{(1)}_{\calC(1)}\right)^{-1}
\bPhi^{(1)} \; .
\end{equation}
2. Application on the result obtained in point 1.,
transformation ${\bF}_{(2)}$
with the data transformed by the transformation ${\bF}_{(1)}$
as well
\begin{equation}
{\bF}_{(2)} ({\bF}_{(1)}(\bx) ) =
{\bF}_{(1)}(\bx) -
{\bF}_{(1)} (\bX_{\calC(2)} )
\left( {\bF}_{(1)}   ( \bPhi^{(2)}_{\calC(2)}  )\right)^{-1}
{\bF}_{(1)} ( \bPhi^{(2)} )  \; ,
\end{equation}
where
\begin{eqnarray}
{\bF}_{(1)} (\bX_{\calC(2)} ) & = &
\bX_{\calC(2)}  - \bX_{\calC(1)} \left( \bPhi^{(1)}_{\calC(1)}\right)^{-1}
\bPhi^{(1)}_{\calC(2)} \label{eq:F1X2} \\
{\bF}_{(1)} (\bPhi^{(2)} ) & = &
\bPhi^{(2)} -  \bPhi^{(2)}_{\calC(1)}  \left( \bPhi^{(1)}_{\calC(1)}\right)^{-1}
\bPhi^{(1)} \; , \\
{\bF}_{(1)} ( \bPhi^{(2)}_{\calC(2)}  ) & = &
\bPhi^{(2)}_{\calC(2)} - \bPhi^{(2)}_{\calC(1)} \left( \bPhi^{(1)}_{\calC(1)}\right)^{-1}
\bPhi^{(1)}_{\calC(2)} \; \label{eq:F1P22}.
\end{eqnarray}
\end{Theo}

Another important feature of the fundamental transformation is that 
its recursive application generates new dimensions of the quadrilateral
lattice. In particular, when applied to the continuous limit of the
quadrilateral lattice -- the
conjugate net -- this procedure explains, on the geometric level, 
the power of using the Darboux type transformations in discretizing the
integrable equations.

\bigskip

\leavevmode\epsfxsize=7cm\epsffile{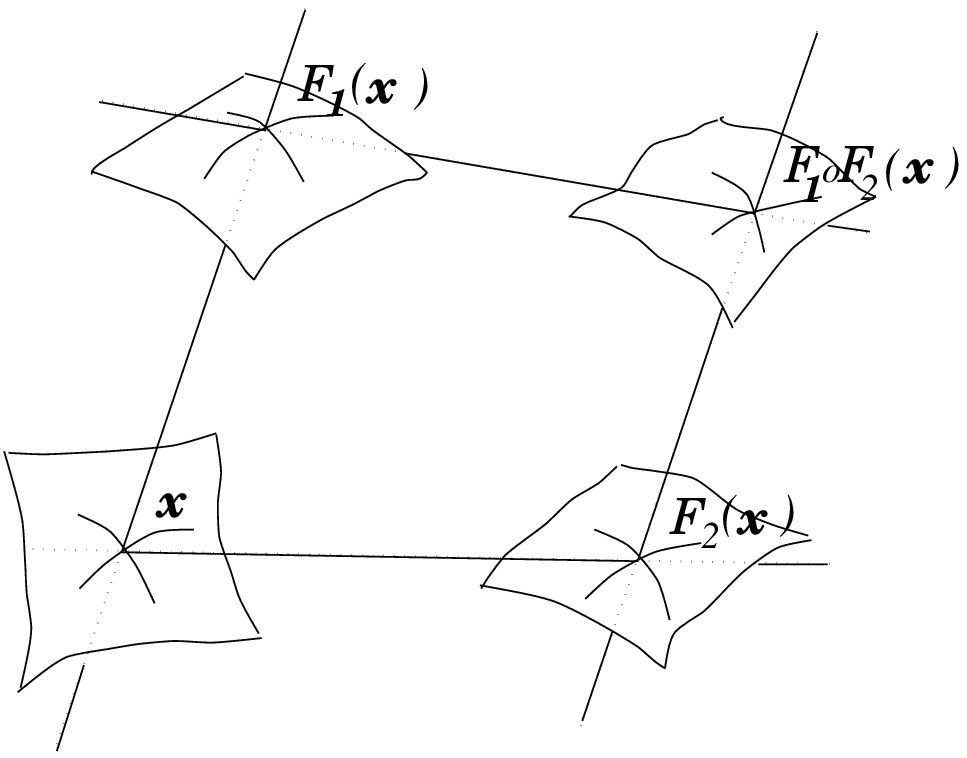}

Figure 7.

\bigskip

\subsection{Vertex operators as transformations of conjugate nets}

The continuous limit of the above theory reduces to the classical theory of
finite transformations of $N$ dimensional conjugate nets. It was recently
shown in~\cite{DMMMS} that these transformations of conjugate nets are
strictly related to the quantum field theoretical formulation of the
multicomponent Kadomtsev--Petviashvilii hierarchy.

The $b$-$c$ system of quantum fields, which appears as the system
of ghost fields in string theory, is constructed in terms of the
anticommutation relations
\begin{eqnarray}
\{b_i(z),c_j(z^\prime)\}&=&\delta_{ij}\delta(z-z^\prime),\\
\{b_i(z),b_j(z^\prime)\}&=&\{c_i(z),c_j(z^\prime)\}=0,
\end{eqnarray}
where $b_i(z)$ and $c_i(z)$, $i=1,\dots ,N$, are free charged
fermion fields defined on the unit circle $S^1$, and
$\delta(z-z^\prime)$ is the Dirac distribution on $S^1$.

The Clifford algebra generated by the $b$-$c$ system admits a
representation in terms of bosonic variables. In this
representation the fields act on the Fock space $\cal F$ of
complex-valued functions
\[
\tau=\sum_\bell\tau(\bell,\bt)\blambda^\bell,
\]
with
\begin{eqnarray}
\bell:=(\ell_1,\cdots,\ell_N)\in\ZZ^N, && \nonumber \\
\bt:=(\bt_1,\dots,\bt_N)\in\CC^{N\cdot\infty},&\quad&
\bt_i:=(t_{i,1},t_{i,2},\dots)\in\CC^\infty, \nonumber \\
\blambda:=(\lambda_1,\dots,\lambda_N)\in\CC^N,&\quad&
\blambda^\bell:=
\lambda_1^{\ell_1}\dots \lambda_N^{\ell_N}
\end{eqnarray}

The representation of the $b$-$c$ generators takes the form
  \cite{Kyoto}:
  
\begin{eqnarray}
b_i(z)\tau(\bell,\bt)&=&(-1)^{\sum_{j>i}\ell_j} z^{\ell_i-1}
\exp(\xi(z,\bt_i))
\tau(\bell-\be_i,\bt-\left[ 1/z \right]\be_i),\\
c_i(z)\tau(\bell,\bt)&=&(-1)^{\sum_{j>i}\ell_j}
z^{-\ell_i}\exp(-\xi(z,\bt_i))
\tau(\bell+\be_i,\bt+\left[ 1/z \right]\be_i),
\end{eqnarray}
where
\[
\xi(z,\bt_i):=\sum_{n=1}^\infty z^n t_{i,n},
\qquad
\left[ 1/z \right] := \left( \frac{1}{z}, \frac{1}{2z^2}, \frac{1}{3z^3},
\ldots \right),
\]
and $\{\be_i\}_{i=1}^N$ are the canonical generators of $\CC^N$.

The Fock space decomposes into a direct sum of charge sectors
\[
{\cal F}=\bigoplus_{q\in\ZZ}{\cal F}_q, \quad {\cal F}_q=\{\tau\in
{\cal F}: Q\tau=q\tau\},
\]
where the total charge operator $Q:=\sum_{i=1}^NQ_i$ is the sum of
$N$ commuting partial charges $Q_i=\lambda_i\partial /\partial\lambda_i$,
$i=1,\dots, N$; they correspond to the  $N$ different flavours of
fermions of the model.

The $N$-component KP hierarchy can be formulated as the following
bilinear identity
\begin{equation}\label{bilineal}
{\cal B}(\tau\otimes\tau)=0,\quad \tau\in{\cal F}_0,
\end{equation}
where
\[
{\cal B}:=\int_{S^1}\frac{d z}{z}\sum_{i=1}^N b_i(z)\otimes
c_i(z).
\]
In terms of the components $\tau(\bell,\bt)$ we have
\begin{eqnarray}\label{bi1}
\int_{S^1} d z\sum_{m=1}^N(-1)^{\sum_{j>m}\ell_j+\ell_j^\prime}
\exp(\xi(z,\bt_m-\bt_m^\prime))z^{\ell_m-\ell_m^\prime-2}&& \\
\times\tau(\bell-\be_m,\bt-\left[ 1/z \right]\be_m)
\tau(\bell^\prime+\be_m,\bt^\prime+\left[ 1/z \right]\be_m)=0 && \nonumber,
\end{eqnarray}
for any $\bt$, $\bt'$ and $\bell$, $\bell'$ such that
$\ell_1+\dots+\ell_N-1=\ell'_1+\dots+\ell'_N+1=0$.

Denote by
\begin{equation} \label{l-fix}
\tau_{ij}(\bell,\bt) = {\cal S}_{ij}\tau(\bell,\bt)
 := \tau(\bell+\be_i-\be_j,\bt);
\end{equation}
the shift operators ${\cal S}_{ij}$ correspond to the so called Schlesinger
transformations in monodromy theory~\cite{Kyoto,JvL} 
and do not alter the neutral character of the
assembly of fermions. Moreover, ${\cal S}_{ij}$ are obvious
symmetries of (\ref{bilineal}).

The $N\times N$ matrix Baker function $\psi$ and its adjoint
$\psi^*$ can be defined in terms of the $\tau$ function as
\begin{eqnarray}\label{baker1}
\psi_{ij}(z,\bt)&=&\varepsilon_{ij}z^{\delta_{ij}-1}
\frac{\tau_{{ij}}(\bt-\left[ 1/z \right]\be_j)}{\tau(\bt)}\exp(\xi(z,\bt_j))
,\\ 
\psi_{ij}^*(z,\bt)&=&\varepsilon_{ji}z^{\delta_{ij}-1}\exp(-\xi(z,\bt_i))
\frac{\tau_{{ij}}(\bt+\left[ 1/z \right]\be_i)}{\tau(\bt)},
\end{eqnarray}
where $\varepsilon_{ij}:={\rm sgn}(j-i)$, $j\neq i$
($\varepsilon_{ii}:=1$).

The connection of the $N$ component hierarchy with the conjugate nets is
given in the following result~\cite{DMMMS}

\begin{Prop} \label{cor:mKP-nets}
The solutions of the N-component KP hierarchy describe $N$
dimensional conjugate nets with coordinates $u_i = t_{i,1}$,
$i=1,\dots,N$, while the remaning times $t_{i,k}$, for $k>1$,
describe integrable iso-conjugate deformations of the nets.

In particular, the rotation coefficients are given by
\begin{equation}
\beta_{ij}(\bt) =\varepsilon_{ij}\frac{\tau_{{ij}}(\bt)}{\tau(\bt)},
\quad
i\neq j,\quad i,j=1,\dots,N,
\end{equation}
the
tangent vectors $\bX_i$ 
are constructed from the rows of the
matrix
\begin{equation} \label{X-psi}
X(\bt)=\int_{S^1}d
z\;\psi(z,\bt)f(z),
\end{equation}
for some distribution  matrix $f(z)\in M_{N\times M}(\CC)$;
the Lam\'e coefficients are
given by the entries of the row matrix
\begin{equation} \label{H-psi}
H(\bt)=\int_{S^1}d z\;g(z)\psi^*(z,\bt),
\quad i=1,\cdots, N,
\end{equation}
for some distribution row matrix $g(z)\in\CC^N$.
\end{Prop}

As one can expect, various elements of the KP theory have their
conterparts on the level of the conjugate nets. 

\begin{Theo}
i) The Schlesinger transformation ${\cal S}_{ij}$ gives rise to the Laplace
transformation $\cL_{ij}$ of the conjugate nets.\\
ii) The action of the (vertex) operators $b_i(p)$ and $c_i(q)$ gives rise to the
L\'evy and the adjoint L\'evy transformations $\cL_i$ and $\cL_i^*$.
Correspondingly\\
iii) the soliton vertex operator generated by the superposition
$b_i(p)c_i(q)$ gives rise to the fundamental transformation.
\end{Theo}

\section{Quadrilateral Hyperplane Lattices}
\label{sec:hyp}
In this Section we introduce and study the properties of quadrilateral 
lattices in the dual space (hyperplane lattices)~\cite{DS4}.  
These considerations turn out to be relevant in the reduction theory
of the quadrilateral lattices, when the introduction of the additional 
geometric structure will
allow to establish a direct connection between point lattices and 
hyperplane lattices.

Generic hyperplanes can
be represented by the co-vectors ${\bf a}^*\in(\RR^M)^*$ and
consist of points ${\bx}$ satisfying the equation
\begin{equation} \label{eq:hyp}
\langle {\bf a}^* | {\bx} \rangle  = a_1^* x^1 + 
\dots + a_M^* x^M  = 1
\; ;
\end{equation}
the equation of the hyperplane passing through $\bf 0$ and
parallel to that represented by ${\bf a}^*$
can be written as $\langle {\bf a}^* | \bx \rangle = 0$ (the co-vector 
${\bf 0}^*$ represents the hyperplane at infinity).

The linear mappings between points and hyperplanes are called 
{\it correlations}. 
Given the sphere of radius $R$ centered at the orgin,
it defines the
corresponding corelation $P$ (called {\it polarity} with respect to that
sphere).
The image $P({\bf v})$ of a point ${\bf v}\in\RR^M$ is called 
the {\it polar hyperplane} of ${\bf v}$, and consists of points $\bx$ 
satisfying equation $ {\bf v}^T \bx  = R^2$, i.e, 
$P({\bf v})= \frac{{\bf v}^T}{R^2}$.

The notions dual to the parallelism of two lines and to the planarity of 
four 
points are contained in the following   

\begin{Def} \label{def:co-parallelity}
i) Two subspaces of co-dimension $2$ are called "co-parallel" if there exists 
a hyperplane passing through the origin and containing them.

\noindent
ii) Four hyperplanes are "co-planar" if the fourth hyperplane contains the 
intersection of the first three.
\end{Def}

A {\it hyperplane lattice} is simply a lattice $\by^*:\ZZ^N \ra (\RR^M)^*$, 
$N\leq M$, in the space of hyperplanes. From Definition 
\ref{def:co-parallelity} we obtain the 
following geometric object dual to the quadrilateral lattice.

\begin{Def} 
The hyperplane lattice $\by^*:\ZZ^N\to (\RR^M)^*$ is "quadrilateral" 
("co-planar") if, for any $i,j=1,\dots,N$, $i\ne j$, the hyperplane 
$T_iT_j \by^*$ contains the subspace $ \by^* \cap T_i\by^* \cap T_j\by^* $.  
\end{Def}
This property reduces again to the Laplace equation:
\begin{equation} \label{eq:eq:h-lat-Lapl-aff}
\D_i\D_j\by^* = (T_iA_{ij}^*)\D_i\by^* + (T_jA_{ji}^*)\D_j\by^*
\; , \quad i\ne j \;.
\end{equation}

A convenient way to construct quadrilateral hyperplane lattices 
makes use of systems of parallel quadrilateral point lattices. We recall that 
two quadrilateral lattices $\bx$, $\bx'$ are parallel (Comberscure related) 
iff $\D_i\bx' \sim \D_i\bx$; in this case the scaled tangent vectors 
of the two lattices can be chosen to be equal, then   
the rotation coefficients $Q_{ij},~Q'_{ij}$ of both lattices coincide and 
the Lam\'e coefficients $H_i$ and $H_i'$ are solutions of the same equation.

\begin{Def}
Consider a system of $M$ parallel and linearly independent point lattices  
$\bx_{(k)}$, $k=1,\dots,M$. Denote by $\by^*_{(k)}$ the 
hyperplane lattice made out of the hyperplanes passing 
through the (corresponding) points of $\bx_{(k)}$ and spanned
by the vectors $\bx_{(l)}$, $l\ne k$, i.e.,
\begin{equation}
\langle \by^*_{(k)} | \bx_{(l)} \rangle = \delta_{kl} \; .
\end{equation}
We call the system $\{\by^*_{(k)}\}$ of hyperplane lattices the 
"dual system" to the system of parallel point lattices $\{\bx_{(k)}\}$.
\end{Def}

If we  arrange the parallel system $\bx_{(k)}$, $k=1,\dots,M$ of point 
lattices as columns of the {\em matrix $\bOm$ of the system}:
\begin{equation}
\bOm = (\bx_{(1)},\dots ,\bx_{(M)} ) \; ,
\end{equation}
then we have the following results.
\begin{enumerate}
\item The dual system $\{\by^*_{(k)}\}$ consists of co-parallel quadrilateral 
hyperplane lattices and is given by the rows of the matrix $\bOm^{-1}$.
\item The rows of the matrix $\bOm$ define a   
system $\{\bx^*_{(k)}\}$ of co-parallel quadrilateral hyperplane lattices, 
which
we call the "adjoint system" to $\bx_{(k)}$,  characterized by the property 
that the forward rotation coefficients  of the system $\{\bx_{(k)}\}$
are the backward rotation coefficients of the system $\{\bx^*_{(k)}\}$:
$Q_{ij}= \tilde{Q}_{ij}^*$.
\end{enumerate}

\section{Reductions of the Quadrilateral Lattice}

We now study some geometrically distinguished reductions of the QL
which possess additional geometric properties that, once imposed
on the initial surfaces, "propagate" everywhere through the
construction of the lattice. Since the quadrilateral lattice
is integrable, these reductions will inherit its integrability
properties.

\subsection{The Symmetric Lattice}

Forward and backward data allow one to introduce the first basic reduction of 
the quadrilateral lattice \cite{DS4}.

\begin{Def} A quadrilateral lattice $\bx$ is {\em symmetric}
iff its forward rotation coefficients are also its backward rotation
coefficients:
\begin{equation} \label{eq:Q=tQ}
\tilde{Q}_{ij} = Q_{ij} \; .
\end{equation}
\end{Def}

The symmetric lattice admits the following different characterizations 
\cite{DS4}.

\begin{enumerate}
\item A quadrilateral lattice is symmetric iff,
for a given set of rotation coefficients $Q_{ij}$,
there exists a $\tau$--function of the lattice such that
\begin{equation} \label{eq:symm}
T_i(\tau Q_{ji})=T_j(\tau Q_{ij}) \; , \quad i\ne j \; ,
\end{equation}
or equivalently, in terms of the corresponding potentials $\rho_i$,
\begin{equation} \label{eq:symm-rho}
\rho_iT_iQ_{ji}=\rho_jT_jQ_{ij} .
\end{equation}

\item A quadrilateral lattice is symmetric iff, 
for a given set of forward tangent vectors $\bX_i$ of the lattice,
there exists a complementary set of the backward tangent vectors 
$\tilde{\bX}_i$ such that the parallelograms
$P(T_i\tilde{\bX}_i , T_j\tilde{\bX}_j)$ and $P(\D_i\bX_j , \D_j\bX_i)$ are
similar.

\item A quadrilateral lattice is symmetric iff, for different indices
$i,j,k$, its rotation
coefficients satisfy the following constraint
\begin{equation}\label{eq:symm-cons-Q}
(T_iQ_{ji})(T_jQ_{kj})(T_kQ_{ik})=(T_jQ_{ij})(T_iQ_{ki})(T_kQ_{jk}).
\end{equation}
\end{enumerate}
We remark that, unlike the others, the characterization (\ref{eq:symm-cons-Q}) 
is {\it local}.


\epsffile{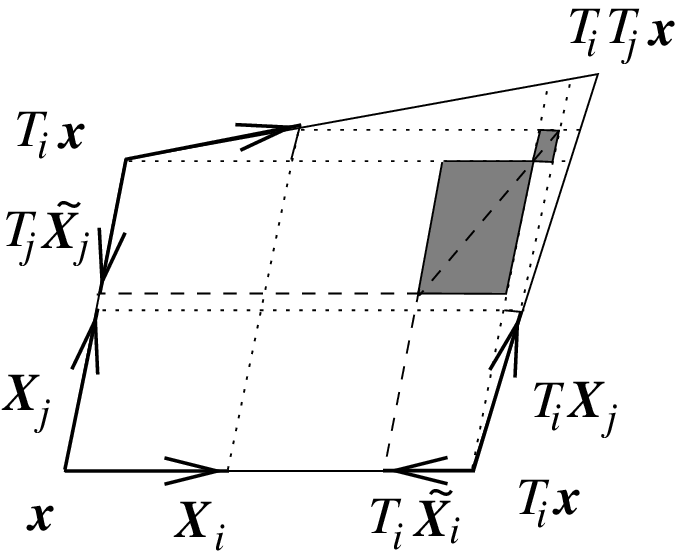}
 
Figure 8.

\bigskip

As we have anticipated, the constraints discussed in this paper allow
one to establish a connection between quadrilateral point lattices
and their duals, the quadrilateral hyperplane lattices. The following
proposition describes this connection in the case of the symmetry constraint.

\begin{Prop} \label{prop:Om-symm}
Given a system of parallel quadrilateral
lattices $\{\bx_{(k)}\}_{k=1}^M$ and the associated matrix $\bOm$, then the
following properties are equivalent.\\ i) The matrix $\bOm$ of the
system is symmetric:
\begin{equation}\label{eq:Omegasymm}
\bOm=\bOm^T.
\end{equation}
ii) The polar hyperplane $P({\bx_{(k)}})$ of the point lattice
$\bx_{(k)}$ coincides with the hyperplane lattice $\bx^*_{(k)}$:
\begin{equation}
P({\bx_{(k)}})={\bx^*_{(k)}} \; ,\quad k=1,..,M \; .
\end{equation}
iii) The lattices $\bx_{(k)}$, $k=1,\dots,M$ are symmetric.
Furthermore the associated tangent vectors $\bX_i$ and $\bX_i^*$
are related in the following way
\begin{equation}\label{eq:symmX}
\bX_i^T = \rho_i(T_i\bX^*_i) , \quad i=1,\dots ,N.
\end{equation}
\end{Prop}

\begin{Cor}
A quadrilateral lattice $\bx$ is symmetric iff it is adjoint to its own
polar.
\end{Cor}

In the continuous limit (\ref{eq:limit}) the symmetric
quadrilateral lattice reduces to a {\em symmetric conjugate net},
for which the rotation coefficients $\beta_{ij}$ satisfying the
Darboux equations (\ref{eq:Darboux}) are symmetric
\begin{equation}  \label{eq:symm-cont}
\beta_{ij} = \beta_{ji} \; .
\end{equation}

\subsection{The Circular Lattice}

The discrete analogue of an $N$ - dimensional orthogonal system of
coordinates is the circular lattice \cite{CDS}.

\begin{Def}
An $N$ - dimensional lattice is circular if and only if
its elementary quadrilaterals are inscribed in circles.
\end{Def}

This notion was first proposed in \cite{2dcl1,2dcl2}
for $N=2$, $M=3$, as a discrete analogue of surfaces parametrized
by curvature lines (see also~\cite{BP2}); later in \cite{Bobenko}  for
$N=M=3$ and, finally, for arbitrary $N\leq M$ in \cite{CDS}; subsequently
a convenient set of
equations characterizing the circular lattices in ${\bf E}^3$ was found in 
\cite{KoSchief2}. A geometric proof of the integrability of the circular
lattice was first given in~\cite{CDS} while the analytic proof of its
integrability was given in~\cite{DMS} through the $\bar\partial$
method.

An elementary characterization of circular quadrilaterals states
that, if a circular quadrilateral is convex, then the sum of its
opposite angles is $\pi$; when the quadrilateral is skew, then its
opposite angles are equal. Other convenient characterizations are the 
following.

\begin{enumerate}
\item A quadrilateral lattice is circular if and only if:
\begin{equation}
\cos \angle (\bX_i , T_i\bX_j) + \cos \angle
(\bX_j , T_j\bX_i) = 0 \; ,
\end{equation}
\item A quadrilateral lattice is circular if and only if \cite{DMS}: 
\begin{equation} \label{eq:circularity1}
\bX_i\cdot T_i\bX_j+\bX_j\cdot T_j\bX_i =0 \; , \quad i\neq j \; .
\end{equation}
\item A quadrilateral lattice $\bx$ is circular iff the scalars
\begin{equation}
v_i:= (T_i \bx + \bx)\cdot \bX_i \; , \quad i=1,\dots ,N
\end{equation}
solve the linear system~(\ref{eq:lin-X}) or, equivalently, iff the
function $|\bx|^2$ (the square of the norm of $\bx$) satisfies
the Laplace equation~(\ref{eq:Laplace}) of $\bx$. This  characterization 
was found in \cite{KoSchief2} and explained geometrically in \cite{AD:q-red}.
\end{enumerate}
 
\bigskip

\leavevmode\epsfxsize=10cm\epsffile{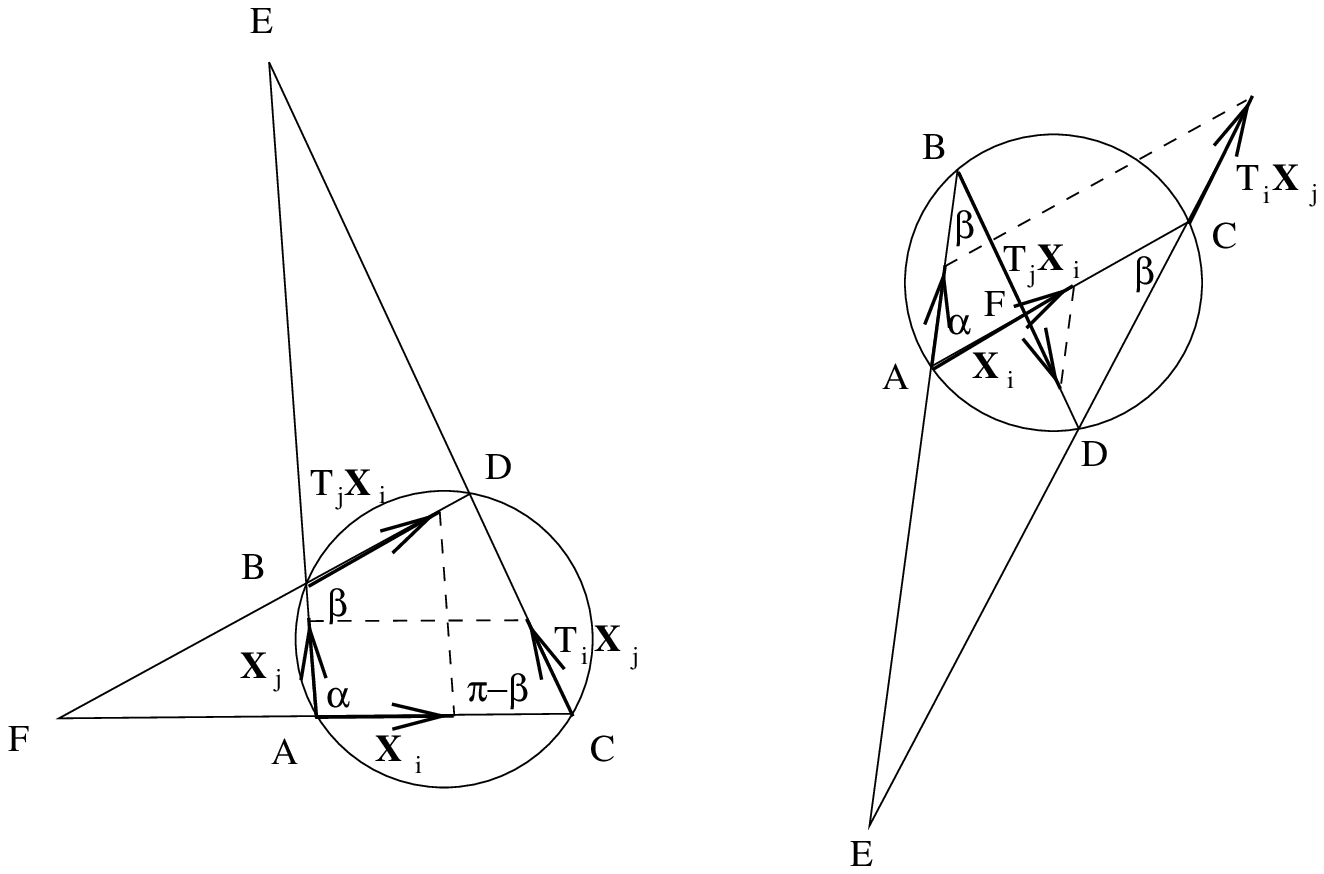}
 
Figure 9.

\bigskip

In addition, the circularity constraint (\ref{eq:circularity1}) implies the 
following formula \cite{DMS}:
\begin{equation} \label{eq:circularity2}
{T_i|\bX_j|^2\over |\bX_j|^2}=1-(T_iQ_{ji})(T_iQ_{ji})  \; ;
\end{equation}
which, compared with equations
(\ref{eq:rho-constr})-(\ref{eq:tau}), allows to fix, without loss
of generality, the backward formulation of the circular lattice in
the following way:
\begin{equation} \label{eq:circularity3}
|\bX_i|^2=  \rho_i=  {T_i\tau \over \tau} \quad \Rightarrow \quad
|T_i\tilde{\bX}_i|^2 =  1/\rho_i = \frac{\tau}{T_i\tau}.
\end{equation}

A distinguished sub-class of circular lattices corresponds to the 
particular case in
which the lattice points $\bx$ belong to the sphere of radius $R$:
$|{\bx}|=R$.
In this case there exists, like for the symmetric reduction, an elegant
relation between point lattices and hyperplane lattices \cite{DS4}.

\begin{Prop}  \label{prop:Om-circ}
Given a system of parallel quadrilateral
lattices $\{\bx_{(k)}\}_{k=1}^M$ and the associated matrix $\bOm$
of the system, then the following properties are equivalent.
\\ i) The matrix $\bOm /R$ is orthogonal:
\begin{equation} \label{eq:Omegaorth}
\bOm \: \bOm^T = \bOm ^T \bOm=R^2 I, \quad \bOm^T=R^2 \bOm^{-1}.
\end{equation}
ii) The polar hyperplane $P({\bx}_{(k)})$ coincides with the dual
hyperplane $R^2 \by^*_{(k)}$:
\begin{equation}
P({\bx}_{(k)})=R^2\by^*_{(k)},\;\;\;k=1,..,M.
\end{equation}
iii) The quadrilateral lattices ${\bx}_{(k)}/R,\;k=1,..,M$ form
an orthonormal basis:
\begin{equation}\label{eq:circ0}
{\bx}_{(i)}\cdot {\bx}_{(j)}=R^2\delta_{ij},\;\;\;i,j=1,..,M.
\end{equation}
In addition, the associated tangent vectors ${\bX}_i,\;{\bX}^*_i$,
$i=1,...,N$, are related by the following formulas
\begin{eqnarray}\label{eq:circ1}
\bX_{i}  =  \frac{\rho_i}{2R^2}T_i( \bOm \bX^{*T}_{i}) & = &
-\frac{\rho_i}{2R^2} \bOm (T_i\bX^{*T}_{i}) , \quad i=1,\dots , N, \\
\label{eq:circ2} T_i\bX^*_{i} & = & -\frac{2}{\rho_i} \bX_i^T \bOm ,
\quad i=1,\dots , N,
\end{eqnarray}
with
\begin{equation}\label{eq:circ3}
|\bX_i|^2= \rho_i,\;\;\;T_i|{\bX}^*_i|^2= \frac{4R^2}{\rho_i}
\end{equation}
and satisfy the circularity constraint (\ref{eq:circularity1}) and
its adjoint
\begin{equation}\label{eq:circularity*}
C^{\circ *}_{ij}:={\bX}^*_i\cdot T^{-1}_i{\bX}^*_j+
{\bX}^*_j\cdot T^{-1}_j{\bX}^*_i=0.
\end{equation}
\end{Prop}

In the continuous limit, equations(\ref{eq:circularity1}) become the
orthogonality conditions
\begin{equation} \label{eq:cont-ort}
\bX_i \cdot \bX_j = 0 , \qquad i\ne j
\end{equation}
and the circular lattice reduces to an orthogonal conjugate net.

\subsection{The Egorov Lattice}

In the previous sections we introduced two integrable reductions
of the quadrilateral lattice: the symmetric and the circular
lattice. Imposing both properties on the initial surfaces of the lattice, 
they 
propagate through the lattice in the unique construction determined by the 
planarity condition and the corresponding
lattice will be an integrable deeper reduction of the
quadrilateral lattice. We call this lattice a {\em Egorov lattice} 
since, in the continuous limit, it reduces to a Egorov system of 
coordinates on
submanifolds~\cite{DS4}.

\begin{Def} \label{Def:Ego}
A Egorov lattice is a circular and symmetric lattice.
\end{Def}

A Egorov lattice can be characterized as follows.

\begin{Prop} 
A quadrilateral lattice is a Egorov lattice iff the
internal angles  corresponding to the
vertices $T_i\bx $ and $T_j\bx $ are right angles; i.e.:
\begin{equation}\label{eq:Ego}
\bX_i\cdot T_i\bX_j=0 ,\;\;i\ne j.
\end{equation}
\end{Prop}
This property was first announced in \cite{Schief-priv} without 
details; it was later shown in \cite{DS4} to be equivalent to Definition 
\ref{Def:Ego}. 
    
If $N=M$, then it turns out that the 
Egorov lattice is a circular lattice invariant 
under translations along the main diagonal \cite{DS4}, i.e., such that:
\begin{equation} \label{eq:d-inv}
T\bX_i = \bX_i \; \Rightarrow T Q_{ij} = Q_{ij}  \; ,
\end{equation}
where $T$ is the total shift: $T:=\prod_{i=1}^{N}T_i$.

We remark that this invariance property implies that the lattice depends
effectively 
on $N-1$ parameters, since it depends on the differences of the variables 
$n_i$:
\begin{equation}
\bX_i = \bX_i(n_1-n_2, n_2 - n_3 , \dots , n_{N-1} - n_N) \; .
\end{equation}

We have been recently informed of a work on the finite gap formulation
of the circular and Egorov lattices \cite{Krich-priv}.

The continuous limits of the above equations:
\begin{equation}
\partial_i \beta_{jk}  = \beta_{ji}\beta_{ik} ,~~~~~ 
\beta_{ij}  = \beta_{ji} ,~~~~~ \bX_i\cdot \bX_j = 0 , \qquad i\ne j  
\end{equation}
characterize submanifolds parametrized by Egorov systems of conjugate
coordinates (Egorov nets). At last, the 
invariance property (\ref{eq:d-inv}) reduces to
\begin{equation}
\sum_{\ell = 1}^N \partial_\ell \beta_{ij} =  0 ,~~
\sum_{\ell = 1}^N \partial_\ell \bX_{i} =  0 ,
\end{equation}
implying that $\beta_{ij} = \beta_{ij}(u_1 - u_2,\dots ,u_{N-1}-u_N)$.
For $N=3$, we recover a classical characterization of the Egorov
net~\cite{Bianchi,DarbouxOS}.

\subsection{Quadratic reductions and the Ribaucour transformation}

Quadrilateral lattices in quadrics generalize (in a sense which will be
explained in Proposition~\ref{prop:st-clat}) the circular lattices. 
Their continuous analogs -- the 
so called {\it quadratic conjugate nets} (see \cite{Eisenhart-TS} and 
references therein) were the subject of intensive study by Ribaucour, 
Darboux, Bianchi
and Eisenhart. The particular type of fundamental transformation
of conjugate nets which preserves the quadratic constraint is called the
Ribaucour transformation. In this section we present the basic geometric
results concerning quadrilateral lattices in quadrics (see \cite{AD:q-red}
for details); in particular, we give the analogue of the Ribaucour 
transformation.

Consider the quadrilateral lattice $\bx$ in the quadric $\cQ$ defined by
the equation  
\begin{equation} \label{eq:quadric}
\bx^T  Q  \bx +  \ba^T \bx + c = 0  \; \; ,
\end{equation}
where $Q$ is a symmetric matrix, $\ba$
is a constant vector, $c$ is a number.

As it was shown in~\cite{AD:q-red}, such a constraint 
is compatible with the geometric integrability scheme of the quadrilateral
lattice.
This integrability statement can be generalized to quadrilateral
lattices in spaces obtained by intersection of many quadric
hypersurfaces like, for example, lattices in the Grassmann manifolds
and in Segr\'e or Veronese varieties. 

To show a simple example, consider a quadrilateral lattice in the standard
$M$ dimensional sphere $S^M$. 
The intersection of the plane of any elementary quadrilateral of the
lattice $\bx:\ZZ^N \ra S^M$
with the sphere $S^M$ is a circle. In the stereographic projection 
$S^M \ra \EE^M\cup \{ \infty \}$ 
circles of the sphere $S^M$ are mapped 
into circles (or straight lines, i.e., circles passing through the infinity 
point) of $\EE^M$, therefore we have:
\begin{Prop} \label{prop:st-clat}
Quadrilateral lattices in the sphere $S^M$ are mapped during the
stereographic projection into multidimensional circular lattices in $\EE^M$;
conversely, any circular lattice in $\EE^M$ can be obtained in this way.
\end{Prop}

Theorem~\ref{th:fund} states that, in order to construct the fundamental
transformation of the lattice $\bx$, we need three new ingredients: 
$\phi$, $\bx_\calC$
and $\phi_\calC$. In looking for the Ribaucour reduction of the 
fundamental transformation,
we can use the following additional information:\\
i) the initial lattice $\bx$ satisfies the quadratic 
constraint~(\ref{eq:quadric}),\\
ii) the final lattice $\cR(\bx)$ should satisfy the same constraint as well.\\
This should allow to reduce the number of the necessary data and, indeed,
to find the Ribaucour transformation, it is enough to know the Combescure
transformation vector $\bx_\calC$ only.
\begin{Prop}
The Ribaucour reduction $\cR(\bx)$ of the fundamental transformation of the
quadrilateral lattice $\bx$ subjected to quadratic constraint~(\ref{eq:quadric})
is determined by the Combescure transformation vector $\bx_\calC$,
provided that $\bx_\calC$ is not annihilated by the bilinear form $Q$ of the constraint
\begin{equation} \label{eq:xCBxC}
\bx_\calC^t  Q \bx_\calC   \ne 0 \; .
\end{equation}
The functions $\phi$ and $\phi_\calC$ entering in formula (\ref{eq:fund})
are then given by
\begin{eqnarray}
\label{def:phi-q}
\phi & = & 2 \bx^t Q \bx_\calC +  \ba^t  \bx_\calC
\; \; , \\ \label{def:phiC-q}
\phi_\calC & =  & \bx_\calC^t Q   \bx_\calC   \; .
\end{eqnarray}
\end{Prop}
As it was shown in~\cite{AD:q-red}, 
the Ribaucour transformation for circular lattices in $\EE^M$, defined 
algebraically
in~\cite{KoSchief2}, can be obtained combining the stereographic projection
with the Ribaucour transformation of the quadrilateral lattices in $S^M$.

To present the Ribaucour reduction of the vectorial fundamental
transformation we use results and notation of Section~2, together with equations 
(\ref{def:phi-q}) and (\ref{def:phiC-q}),
to obtain
\begin{eqnarray}
\label{def:phi-qv}
\phi^k & = &2 \bx^t Q  \bx_{\calC,k}  +
 \ba^t \bx_{\calC,k}
\; \; , \\ \label{def:phiC-qv}
\phi^k_{\calC,k} & = & \bx_{\calC,k}^t  Q \bx_{\calC,k}  \; .
\end{eqnarray}
Equations (\ref{eq:int-phiC}) and (\ref{def:phi-qv}) lead to
\begin{equation}
\D_i \left( \phi^k_{\calC,\ell} + \phi^\ell_{\calC,k} \right) =
(T_i\bx_{\calC,k}^t + \bx_{\calC,k}^t)Q(\D_i\bx_\ell) +
(T_i\bx_{\calC,\ell}^t + \bx_{\calC,\ell}^t)Q(\D_i\bx_k)
\end{equation}
which implies that
\begin{equation} \label{eq:phi-klk}
\phi^k_{\calC,\ell} + \phi^\ell_{\calC,k} =
2 \bx_{\calC,k}^t Q \bx_{\calC,\ell} \; ;
\end{equation}
the constant of integration was found from the condition
$(\cR_k\circ\cR_\ell)(\bx)\subset \cQ$.
\begin{Prop} \label{prop:vect-Rib}
The vectorial Ribaucour transformation ${\bf R}$, i.e., the reduction
of the vectorial fundamental transformation (\ref{eq:vect-fund})
compatible with the
quadratic constraint (\ref{eq:quadric}), is given by the
following constraints
\begin{eqnarray}
\label{eq:bphi-qv}
\bPhi^T & = & 2 \bx^t Q  \bX_\calC   +
 \ba^T \bX_\calC \; \; , \\
\label{eq:bPhiC-qv}
\bPhi_{\calC} + \bPhi_{\calC}^T & = & 2 \bX_{\calC}^T  Q \bX_{\calC}  \; .
\end{eqnarray}
\end{Prop}
One can also consider the superposition of several 
Ribaucour transformations and
show the corresponding permutability theorem.
\begin{Theo} \label{prop:sup-vect-Rib}
Assume the following splitting of the data of the vectorial Ribaucour
transformation
\begin{equation} \label{eq:split-Rib}
\bPhi=\left( \begin{array}{c} \bPhi^{(1)} \\ \bPhi^{(2)} \end{array}
\right) \; , \quad
\bX_\calC = \left( \bX_{\calC(1)}, \bX_{\calC(2)} \right) \; , \quad
\bPhi_\calC = \left( \begin{array}{cc}  
\bPhi^{(1)}_{\calC(1)} &  \bPhi^{(1)}_{\calC(2)} \\
\bPhi^{(2)}_{\calC(1)} &  \bPhi^{(2)}_{\calC(2)}  \end{array}
\right)  \; ,
\end{equation}
associated with the partition $K=K_1+K_2$. Then the vectorial Ribaucour
transformation ${\bf R}(\bx)$ is equivalent to the following
superposition of vectorial Ribaucour transformations:\\
1. Transformation ${\bf R}_{(1)}(\bx)$ with the data $\bPhi^{(1)}$,
$\bX_{\calC(1)}$, $\bPhi^{(1)}_{\calC(1)}$.\\
2. Application on the result obtained in point 1. the transformation 
${\bf R}_{(2)}$
with the data
${\bf R}_{(1)} (\bX_{\calC(2)} )$,
${\bf R}_{(1)}   ( \bPhi^{(2)}_{\calC(2)}  )$,
${\bf R}_{(1)} ( \bPhi^{(2)} )$ given by $\cR$-analogs of
formulas~(\ref{eq:F1X2})--(\ref{eq:F1P22}).
\end{Theo}
Combining the stereographic projection
with the vectorial Ribaucour transformation of the quadrilateral lattices in 
$S^M$, one can obtain the vectorial Ribaucour transformation for 
circular lattices in $\EE^M$
and prove its permutability property. This result was also found 
independently in~\cite{LiuManas-SR}, directly on the level of circular
lattices in $\EE^M$.

\end{document}